\documentclass[twocolumn,showpacs,10pt]{revtex4}
\usepackage{graphicx}
\usepackage{amsmath,amsfonts,amssymb,bm}

\begin{document}

\title{Stochastic equation for a jumping process with long-time correlations}

\author
{T. Srokowski and A. Kami\'nska}

\affiliation{
 Institute of Nuclear Physics, PL -- 31-342 Krak\'ow, Poland }

\date{\today}

\begin{abstract}
A jumping process, defined in terms of jump size distribution and waiting time
distribution, is presented. The jumping rate depends on the process value.
The process, which is Markovian and stationary, relaxes to an equilibrium and
is characterized by the power-law autocorrelation function. Therefore, it can
serve as a model of the $1/f$ noise as well as a model of the stochastic force
in the generalized Langevin equation. This equation is solved for the noise
correlations $\sim 1/t$; the resulting velocity
distribution has sharply falling tails. The system preserves
the memory about the initial condition for a very long time.

\end{abstract}

\pacs{05.40.-a, 02.50.-r, 05.10.Gg}

\maketitle

\section{Introduction}

A jumping process can be defined in terms of two probability distributions
which determine a jump size and a waiting time between consecutive jumps.
One usually assumes that both distributions are independent of each other.
Such process is often regarded as a generalized form of the random walk 
and used to describe a diffusive transport. That approach,
known as the continuous-time random walk theory \cite{mon}, is able to account 
for various forms of diffusion, both normal and anomalous, by a suitable 
choice of the probability distributions defining the process \cite{met}.
Power-law dependences are especially interesting
\cite{mon1,mon2}. A stochastic trajectory characterized by jump sizes
so distributed, exhibits a pattern
typical for the L\'evy flights and features systems which reveal
the enhanced diffusion \cite{bou}. On the other hand, long
tails of the waiting time distribution (long rests) evoke the opposite effect:
they are responsible for the subdiffusion \cite{met,yus}.
Processes which possess such tails are often treated in terms of the fractional
diffusion equation \cite{bal,bar,met1,sca}.

For uniform distribution of jumps in time, i.e. if the waiting time probability
density has the exponential form, the jumping process relaxes to some 
stationary equilibrium. The kangaroo process \cite{fri} provides a simple 
and well-known example. Instead of the jump size distribution, this
process assumes a probability distribution of the process value after the
jump and, in addition, a jumping rate which depends on the process value.
An advantage of the kangaroo process from the point of view of possible 
applications stems from the fact that it can be easily constructed 
for arbitrary correlations. A need for models of correlated noises is obvious. 
For example, the long correlations, both in space and time, 
arise as a result of the fast modes removal
procedure \cite{med,yak,han}. Long tails of the correlation function emerge
also in the relaxation process of a system coupled to a fractal heat bath
via a random matrix interaction \cite{lut}. In those cases the stochastic
dynamics obeys the generalized, non-Markovian, Langevin equation
and the Monte Carlo simulation of solutions
requires a specific model of the noise. Unfortunately, the kangaroo process
is not suitable to model noises with power law correlations: 
the distribution of the stochastic variable during the trajectory evolution 
is biased because the waiting time distribution changes its shape when it
is inserted into the generalized Langevin equation \cite{sro}. As a result,
the relaxation to the thermal equilibrium cannot be achieved.

In this paper we consider a simple power-law correlated jumping process
which is exempt of that difficulty. Our objective is not only to analyze 
the master equation for that process
but above all to obtain the stochastic variable itself
by solving a stochastic equation. Therefore the presented procedure can be
utilized as a noise model for numerical simulations of the stochastic
trajectories in the framework of the Langevin formalism.
We define the process and discuss corresponding equations in Sec.II.
The expression for the autocorrelation function is derived in Sec.III,
whereas Sec.IV is devoted to the application of the process as a model of some
specific form of the correlated noise in the generalized Langevin equation. The
main results are summarized in Sec.IV.

\section{Definition of the process}

We assume that a stochastic process is step-wise,
i.e. a process value ${\bf x}$ is constant within the time intervals
$(t_i, t_{i+1})$: ${\bf x}(t)={\bf x}_i$ for $t\in(t_i, t_{i+1})$. Jumping 
times $t_i$ are randomly distributed and jumping rate $\nu({\bf x})$ depends 
on the process value. The size of the jump, defined as the difference between 
the values of ${\bf x}$ after and before the jump, is determined from 
a given probability distribution $Q(\delta{\bf x})$.
Then the stochastic trajectory ${\bf x}(t)$ obeys the following equation
\begin{equation}
\label{stoch}
{\bf x_{i+1}}={\bf x_i}+\delta{\bf x}
\end{equation}
where the waiting time $\tau=t_{i+1}-t_i$ is governed by the Poissonian
distribution:
\begin{equation}
\label{poi}
P_P(\tau)=\nu({\bf x}){\mbox e}^{-\nu({\bf x})\tau}
\end{equation}
which determines the probability density that a jump occurs in the interval
$(\tau,\tau+d\tau)$. The initial condition for the Eq.(\ref{stoch}),
${\bf x}(t_0)={\bf x}_0$, follows from a given probability
distribution $P_0({\bf x})$. The Eq.(\ref{stoch}) is stochastic
because it determines the time evolution
of the stochastic variable ${\bf x}$, in contrast to the master equation
which can give us only probability distributions. The trajectory ${\bf x}(t)$
can be constructed step by step by sampling consecutive values of
$\tau$ and $\delta{\bf x}$ from the distributions $P_P$ and $Q$,
respectively.

The process is Markovian and stationary. The transition probability
$p_{tr} d{\bf x}$ that the process value is between ${\bf x}$ and
${\bf x}+d{\bf x}$ at an infinitesimal time $\Delta t$,
providing it was equal ${\bf x}'$ at $t=0$, is given by:
\begin{equation}
\label{trkp}
\begin{split}
p_{tr}({\bf x},\Delta t|,{\bf x}',0)&=
 \{1-\nu({\bf x}') \Delta t\} \delta({\bf x}'-{\bf x})+\\
+& \nu({\bf x}') \Delta t Q({\bf x}-{\bf x}').
\end{split}
\end{equation}
In the above expression we have utilized the fact that $p_{tr}$ may
depend only on time differences. The first term on the right-hand side 
of Eq.(\ref{trkp}) is the probability that no jump occurred in the time 
interval $(0,\Delta t)$. The term $\nu({\bf x}') \Delta t$ means the
probability that one jump occurred. The master equation for a probability
density $p({\bf x},t)$ can be obtained by calculating the time derivative
from $p({\bf x},t)$ and taking into account all possible
initial values ${\bf x}'$:
\begin{equation}
\label{2ks}
\begin{split}
&\frac{\partial}{\partial t}p({\bf x},t) =\\
=&\lim_{\Delta t\to 0^+}\left[ \int p_{tr} ({\bf x},\Delta t \mid
{\bf x}',0) p({\bf x}',t) d{\bf x}' -p({\bf x},t)\right]/\Delta t.
\end{split}
\end{equation}
We get the master equation in the following form
  \begin{equation}
  \label{fpkp}
  \frac{\partial}{\partial t}p({\bf x},t) = -\nu({\bf x})p({\bf x},t) +
  \int Q({\bf x}-{\bf x}')\nu ({\bf x}') p({\bf x}',t) d{\bf x}'.
  \end{equation}

The jumping process described above is still too general and thus we introduce
additional restrictions. Let ${\bf x}$ be a two-dimensional vector,
${\bf x}=(x_1,x_2)$, with the unit length: $|{\bf x}|=1$. Therefore we require
the norm to be preserved during the jumps. 
With these assumptions, the process can
be described in terms of a single angle variable $\phi$: $x_1=\cos(\phi)$ and
$x_2=\sin(\phi)$. For the probability density $Q$ we take the Gaussian:
  \begin{equation}
  \label{ker}
Q(\delta{\bf x})\sim {\mbox e}^{-({\bf x}-{\bf x}')^2/2\sigma^2}=
N{\mbox e}^{\cos(\phi-\phi')/\sigma^2},
  \end{equation}
where $\sigma$ is a given width and the normalization constant
$N=1/[2\pi I_0(1/\sigma^2)]$
contains the modified Bessel function. We will demonstrate in the following
that the jumping rate $\nu$ determines the shape of the
process autocorrelation function. It appears algebraic if $\nu$ is given by the
expression
  \begin{equation}
  \label{nu}
\nu(\phi)=\frac{4}{1-\alpha}\frac{|\sin(\phi)|^\alpha}{|\cos(\phi)|}
  \end{equation}
where $0<\alpha<1$. Taking into account the above assumptions, we obtain
the master equation Eq.(\ref{fpkp}) in the one-dimensional form:
  \begin{equation}
  \label{mast}
  \frac{\partial}{\partial t}p(\phi,t) = -\nu(\phi)p(\phi,t) +
  \int_0^{2\pi} Q(\phi-\phi')\nu(\phi') p(\phi',t) d\phi'.
  \end{equation}
The equilibrium solution of Eq.(\ref{mast}), $P(\phi)$, has to satisfy 
the condition $\nu(\phi)P(\phi)\!=$const. Therefore, $P(\phi)$ becomes 
quite simple:
  \begin{equation}
  \label{psta}
P(\phi)=1/\nu(\phi).
  \end{equation}
Since for the jumping rate (\ref{nu}) $\int_0^{2\pi}1/\nu(\phi)d\phi=1$, 
$P(\phi)$ is properly normalized.

Numerical simulation of stochastic trajectories requires random numbers
distributed like $Q(\delta{\bf x})$, according to Eq.(\ref{ker}). For that
purpose we apply the rejection method which allows us to avoid evaluating
complicated integrals. The algorithm is the following. First we sample 
uniformly distributed random numbers $\delta\phi=\phi-\phi'$ from the interval
$(0,2\pi)$. Then $q=Q(\delta\phi)$ is calculated and this value 
is compared with another random number,
$r_Q$, uniformly distributed within the interval $(Q_{min},Q_{max})$ where
$Q_{min}$ and $Q_{max}$ denote the minimum and maximum values of $Q$,
respectively. If $r_Q>q$ then $\delta\phi$ is accepted, otherwise it is
rejected and the sampling procedure is repeated.

\section{Autocorrelation function for the jumping process}

The autocorrelation function of the process (ACF),
${\cal C}(t)=\langle{\bf x}(0){\bf x}(t)\rangle$, where the average is taken
over the stationary distribution
$P({\bf x})$, can be evaluated from the following expression \cite{gar}
\begin{equation}
\label{cov1}
{\cal C}(t)=\int\int {\bf x}'(t_0){\bf x}(t_0+t)P({\bf x},t|{\bf
x}')p({\bf x}',t_0)d{\bf x}d{\bf x}'.
 \end{equation}
The conditional probability of passing from ${\bf x}'$ to ${\bf x}$ during the
time $t$, $P({\bf x},t|{\bf x}')$, can be obtained by taking into
account all possible paths leading from ${\bf x}'$ to ${\bf x}$ and summing over
the jumps \cite{kam}. The final formula for the Laplace transform of ACF
can be expressed by the following series
\begin{widetext}
\begin{equation}
\label{cov2}
\begin{split}
\widetilde{\cal
C}(s)&=\int_0^{2\pi}\frac{1}{\nu(\phi)}\frac{1}{\nu(\phi)+s}d\phi+\int_0^{2\pi}
\frac{Q(\phi-\phi')}{\nu(\phi)+s}\frac{\cos(\phi-\phi')}{\nu(\phi')+s}d\phi
d\phi'+\\
+&\sum_{k=2}^\infty\int_0^{2\pi}\frac{\cos(\phi-\phi_0)}{\nu(\phi_0)+s}\frac{Q(\phi-\phi_{k-1})}
{\nu(\phi) + s }
\left[\prod_{i=2}^k\frac{Q(\phi_{i-1}-\phi_{i-2})}
{\nu(\phi_{i-1}) + s
}\nu(\phi_{i-1})d\phi_{i-1}\right]d\phi_0d\phi .
\end{split}
 \end{equation}
 Inverting the Laplace transform we obtain the final expression for ACF:
\begin{equation}
\label{covt}
\begin{split}
{\cal C}(t)=&4\int_0^{\pi/2}\frac{{\mbox e}^{-\nu(\phi)t}}{\nu(\phi)}d\phi+\\
+&8N\int_0^{\pi/2}\int_0^{\pi/2}\left({\mbox
e}^{\cos(\phi-\phi')/\sigma^2} -{\mbox e}^{-\cos
(\phi-\phi')/\sigma^2}\right)\cos(\phi-\phi' ) \frac{{\mbox
e}^{-\nu(\phi')t}-{\mbox
e}^{-\nu(\phi)t}}{\nu(\phi)-\nu(\phi')}d\phi d\phi'+\dots.
\end{split}
\end{equation}
\end{widetext}

We are interested in the asymptotic bahaviour of
${\cal C}(t)$ for large $t$. In this limit the first term of Eq.(\ref{covt})
can be estimated easily.  Because of the
exponential dependence of the integrand on $t$, only a vicinity of $\phi=0$
contribute to the integral: $\nu\lesssim 1/t$.
Therefore the first term can be approximated by the integral
$\int_0^\infty\exp(-\phi^\alpha t)/\phi\; d\phi\sim t^{1-1/\alpha}$. In
the second term we first calculate the integral over $\phi$:
$\int_0^{\pi/2}[\int_0^{\pi/2} d\phi]d\phi'$. If we take the limit of large
$t$ in the inner integral, the exponential containing $\phi'$ can
be dropped. Moreover $\nu(\phi)$ becomes negligible, comparing to
$\nu(\phi')$, as well as $\phi$ in the arguments of the cosine function. Then
for any $\phi'>0$ we have:
$$
\frac{{\mbox
e}^{-\nu(\phi')t}-{\mbox e}^{-\nu(\phi)t}}{\nu(\phi)-\nu(\phi')}\approx
{\mbox e}^{-\phi^\alpha t}/\nu(\phi')
$$
and the integral over $\phi$ can be easily evaluated. The required
time dependence is of the
form $t^{-1/\alpha}$ which means that the second term falls with time faster
than the first one. The same conclusion refers to the higher terms. The
second term has a simple asymptotic dependence also on the kernel width $\sigma$.
Expanding the exponential functions over $1/\sigma$ and taking into account that
$\lim_{x\to 0}I_0(x)=1$, we find that the second term decreases like
$1/\sigma^2$ for large $\sigma$. We finally conclude that the ACF
can be well approximated by the first term of the Eq.(\ref{covt}) and its tail
is algebraic:
\begin{equation}
\label{covt1}
{\cal C}(t)\sim t^{1-1/\alpha}~~~~\mbox {for}~~~~t\to \infty.
\end{equation}

Fig.1 presents ACF for $\alpha=0.5$; ${\cal C}(t)$ was
calculated from the definition, by means of single trajectories evolution
according to Eq.(\ref{stoch}), for $\sigma=1$ and $\sigma=2.5$. The
equilibrium probability distribution $P(\phi)$ was taken as the initial
condition. The result for the larger value of $\sigma$ agrees very well with
the first term in Eq.(\ref{covt}) and it can be parameterized by the function
\begin{equation}
\label{covt2}
{\cal C}(t)=\frac{1-{\mbox e}^{-8t}}{8t}.
\end{equation}
\begin{figure}[htb]
\includegraphics[width=8.5cm]{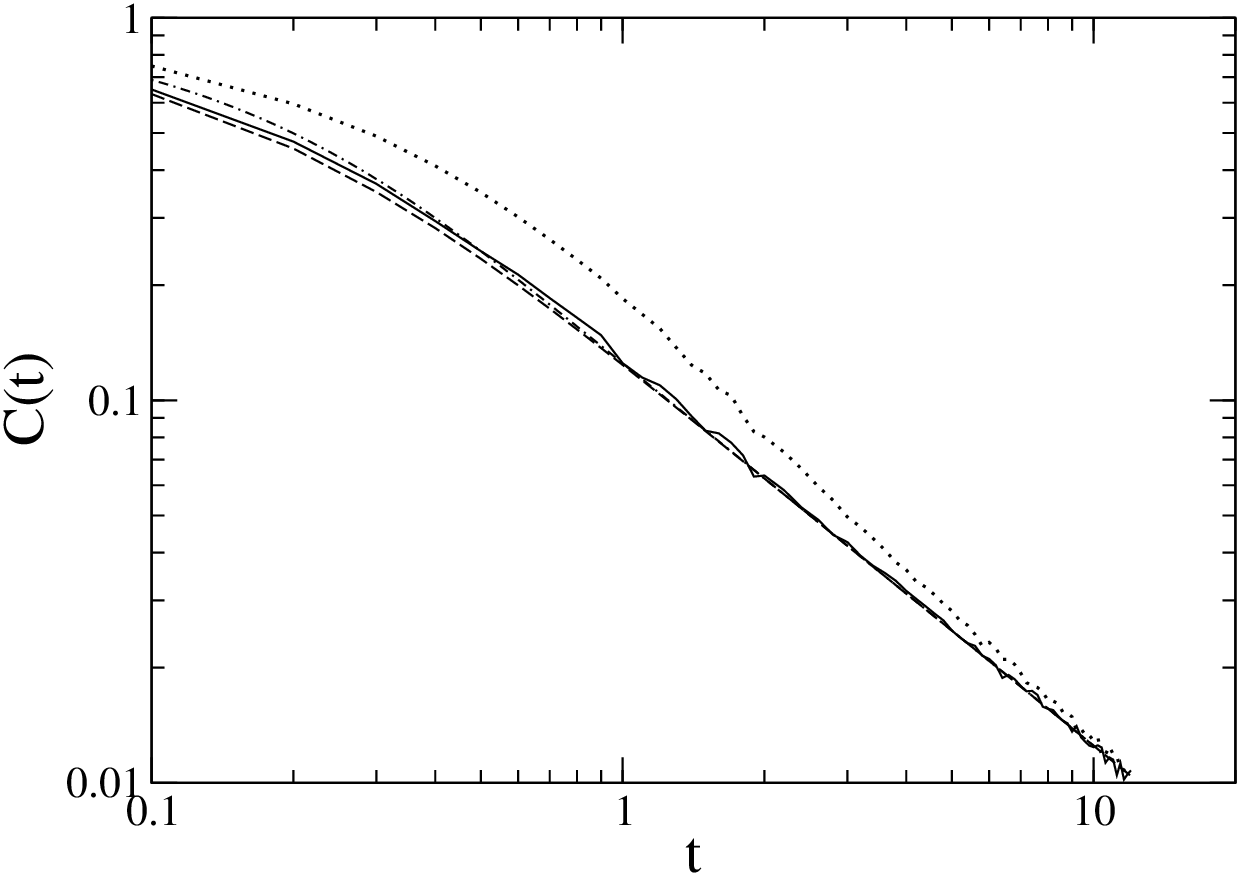}
\caption{Autocorrelation function ${\cal C}(t)$ for the jumping rate $\nu$
given by Eq.(\ref{nu}) with $\alpha=0.5$. Numerical simulations have been
performed for $\sigma=2.5$ (solid line) and $\sigma=1$ (dotted line).
The first term in Eq.(\ref{covt}) is also shown (dashed line),
as well as the parameterization by Eq.(\ref{covt2}) (dashed-dotted line).}
\end{figure}

The existence of long tails of ACF means that the power spectrum 
of the process, defined by the Fourier transform ${\cal F}({\bf x})$ as
$S(\omega)=|{\cal F}({\bf x})|^2$, is strongly enhanced at $\omega=0$. The 
power spectrum can be obtained directly from ${\cal C}(t)$ by using the
Wiener-Khinchin theorem \cite{gar}: $S(\omega)={\cal F}({\cal C}(t))$. For
$0.5<\alpha<1$ we get the following result:
\begin{equation}
\label{pow}
S(\omega)\sim \omega^{-1/\alpha}.
\end{equation}
Then our jumping process is characterized by the algebraic power spectrum
and becomes the $1/f$ noise for $\alpha\to 1$. Overpopulation of small
frequency values is due to the fact that the process is dominated by long
waiting times between consecutive jumps.
Such long intervals correspond to small values of
$\phi$, i.e. to the evolution along the $x_1$ axis. A quantity
$s=1/\nu$, which means the average of the Poissonian distribution (\ref{poi}),
is well suited to characterize long rests. The statistics of $s$ is directly
connected with the process value probability distribution $P(\phi)$ and,
in accordance with that, the
density distribution of $s$ in the equilibrium, $\psi(s)$, can be derived
from the equation $|\psi(s)ds|=|P(\phi)d\phi|$. In the limit of large $s$ we
obtain $\psi(s)\sim s^{-1/\alpha}$ and this result means that
the Poissonian waiting time distribution with variable
jumping rate can possess, effectively, power law tails.

The jumping process with $\alpha=0.5$ resembles a deterministic dynamical
system: the Lorentz gas of periodically distributed hard disks. In this 
lattice a particle is elastically reflected by the discs and wanders 
freely among them. Free paths of the particle are infinite 
at directions parallel to the symmetry axes. The system is
characterized by the velocity autocorrelation function with the tail $1/t$,
analogously to Eq.(\ref{covt2}), and by the power spectrum
$S(\omega)\sim|\ln(\omega)|$. However, the long free path distribution falls
faster then its stochastic counterpart, like $s^{-3}$ \cite{ledous}.

\section{Application to the generalized Langevin equation}

If a Brownian particle is driven by a stochastic force with a finite
correlation time, the time evolution of the velocity obeys
the generalized Langevin equation \cite{mori,lee}:
\begin{equation}
\label{gle}
m \frac{d{\bf v}(t)}{dt} = -\frac{\partial V(\bf r)}{\partial{\bf r}}
-m\int_0^t K(t-\tau){\bf v}(\tau)d\tau + {\bf F}(t)
\end{equation}
where $V({\bf r})$ is a position-dependent external potential, ${\bf F}(t)$ is a
stochastic force and $m$ denotes the mass of the
particle. The integro-differential equation (\ref{gle}) can be solved
numerically for any $V({\bf r})$ if we apply a concrete model for the noise
$F(t)$. In the case $V({\bf r})=0$ the Eq.(\ref{gle}) is manageable by Laplace
transforms. We obtain the following solution:
\begin{equation}
\label{solv}
{\bf v}(t)=R(t){\bf v}(0)+m^{-1}\int_0^t R(t-\tau)\,{\bf F}(\tau)\,d\tau,
\end{equation}
where the Laplace transform of the resolvent $R(t)$ is given by the equation
\begin{equation}
\label{rods}
\widetilde R(s)=1/[s+\widetilde K(s)].
\end{equation}

In the Eq.(\ref{gle}) the usual damping term -- proportional to the velocity --
which appears in the ordinary Langevin equation, has been substituted by the
retarded friction in the form of a memory kernel to ensure proper
characteristics of the equilibrium, namely the equation should satisfy the
second fluctuation-dissipation theorem \cite{kubo}. The kernel $K(t)$ has to be
proportional to the noise ACF ${\cal C}(t)$: $K(t)={\cal C}(t)/mk_BT$ where
$T$ is the temperature which characterizes the heat bath and $k_B$ is
the Boltzmann constant. The introduction of memory friction changes 
the shape of the velocity autocorrelation function considerably: 
it is no longer restricted to the exponentials.

We wish to demonstrate how the process described in the preceded Sections
can be applied as a model for the driving stochastic force in Eq.(\ref{gle}).
For that purpose we choose ACF possessing the tail $\sim 1/t$ which characterizes
e.g. the noise-induced Stark broadening \cite{frisch1} and nuclear collisions
in the framework of a dynamical model \cite{sron}.
It can be also found in problems connected with phenomena in disordered
media \cite{bou}. This form of ACF is of special importance for molecular
dynamics because it corresponds to the problem of scattering inside 
a periodic lattice \cite{zach}. Let us then consider the ACF given by
Eq.(\ref{covt2}). Moreover we assume $\langle{\bf F}(t)\rangle=0$. In this case
$\widetilde K(s)=\ln(1+8/s)/8$ and Eq.(\ref{rods}) reads
\begin{equation}
\label{rods1}
\widetilde R(s)=\frac{1}{s+\ln(1+8/s)/8}.
\end{equation}
In order to obtain the resolvent $R(t)$ we need to inverse the above transform.
Computing the usual contour integral produces the following result
\begin{equation}
\label{res}
\begin{split}
R(t)&={\mbox e}^{-at}\,(c_1\sin bt +c_2\cos bt)+\\
-& {\displaystyle 8\int_0^8 \frac{{\mbox e}^{-tx}\;dx}
{[8x-\ln(8/x-1)]^2+\pi^2}},
\end{split}
\end{equation}
where the constants $a=0.3511$, $b=0.2995$, $c_1=0.2297$ and $c_2=1.603$ follow
from the numerical evaluation of poles in the Eq.(\ref{rods1}). The resolvent
$R$ has the interpretation of the velocity autocorrelation function 
${\cal C}_v$,
\begin{equation}
\label{vaf}
{\cal C}_v(t)=\langle{\bf v}(0){\bf v}(t)\rangle=\frac{k_BT}{m}R(t).
\end{equation}
R(t) falls from $R(0)=1$ to negative values and then rises, approaching
zero very slowly from below. The behaviour of ${\cal C}_v(t)$ at $t\to\infty$
is determined by the integral in Eq.(\ref{res}). In this limit, it becomes
simpler: $\sim \int_0^8 {\mbox e}^{-tx}/\ln^2x\;dx$. Integrating over $t$
yields the integrand in the form ${\mbox e}^{-tx}/(x\ln^2x)$ and the integral
over $x$ can be estimated \cite{kam} as $\sim 1/\ln t$.
The final expression reads
\begin{equation}
\label{vaf1}
{\cal C}_v(t)\sim \frac{-1}{t\ln^2t}~~~~~(t\to \infty).
\end{equation}
Therefore the tail of ${\cal C}_v(t)$ diminishes very slowly, like the
tail of ${\cal C}(t)$, and it is negative.

The velocity autocorrelation function determines the transport properties of
the system: the diffusion coefficient can be expressed in terms of the
Laplace transform of ${\cal C}_v(t)$ in the form:
${\cal D}=\widetilde{\cal C}_v(s=0)$. Since for ${\cal C}(t)$ given by
Eq.(\ref{covt2}) ${\cal D}=0$, the transport is subdiffusive. We come
to the same conclusion by the direct calculation of the position variance
$\langle{\bf r}^2\rangle(t)$. Integrating ${\cal C}_v(t)$ twice over time, we
get the following estimation:
\begin{equation}
\label{dif}
\langle{\bf r}^2\rangle(t)\sim\mbox{li}(t)\approx t/\ln t~~~~~(t\to \infty).
\end{equation}
Therefore the subdiffusion is very weak and hardly distinguishable
from the normal diffusion. The same form of the anomalous diffusion has
been found in a chaotic (deterministic) system \cite{gei} and it has been
attributed to the intermittency.

Our aim is to study the motion of the particle by a direct simulation of
stochastic trajectories, assuming that the driving force in Eq.(\ref{gle})
is modeled by means of the stochastic process ${\bf x}(t)$ and satisfies 
Eq.(\ref{stoch}). We restrict our analysis to the case $V({\bf r})=0$.
Inserting of the solution of Eq.(\ref{stoch}) into Eq.(\ref{solv}) yields the
two-dimensional trajectory of the particle velocity:
\begin{equation}
\label{solv1}
\begin{split}
&{\bf v}(t)=R(t){\bf v}(0)+\\
+&m^{-1}\left[{\bf x}_{n+1}\int_0^{t-t_n} R(\tau)\,d\tau+
\sum_{k=1}^n {\bf
x}_k\int_{t-t_k}^{t-t_{k-1}}R(\tau)\,d\tau\right],
\end{split}
\end{equation}
where by sampling of the consecutive jumping times $t_k$ we apply
Eq.(\ref{poi}) with $\alpha=0.5$.
Moreover, in the following we take the kernel width $\sigma=2.5$.
A simple quantity one can evaluate from the Eq.(\ref{solv1}) is the time
dependence of the velocity variance $\langle{\bf v}^2(t)\rangle$
where the average is taken
\begin{figure}[htb]
\includegraphics[width=8.5cm]{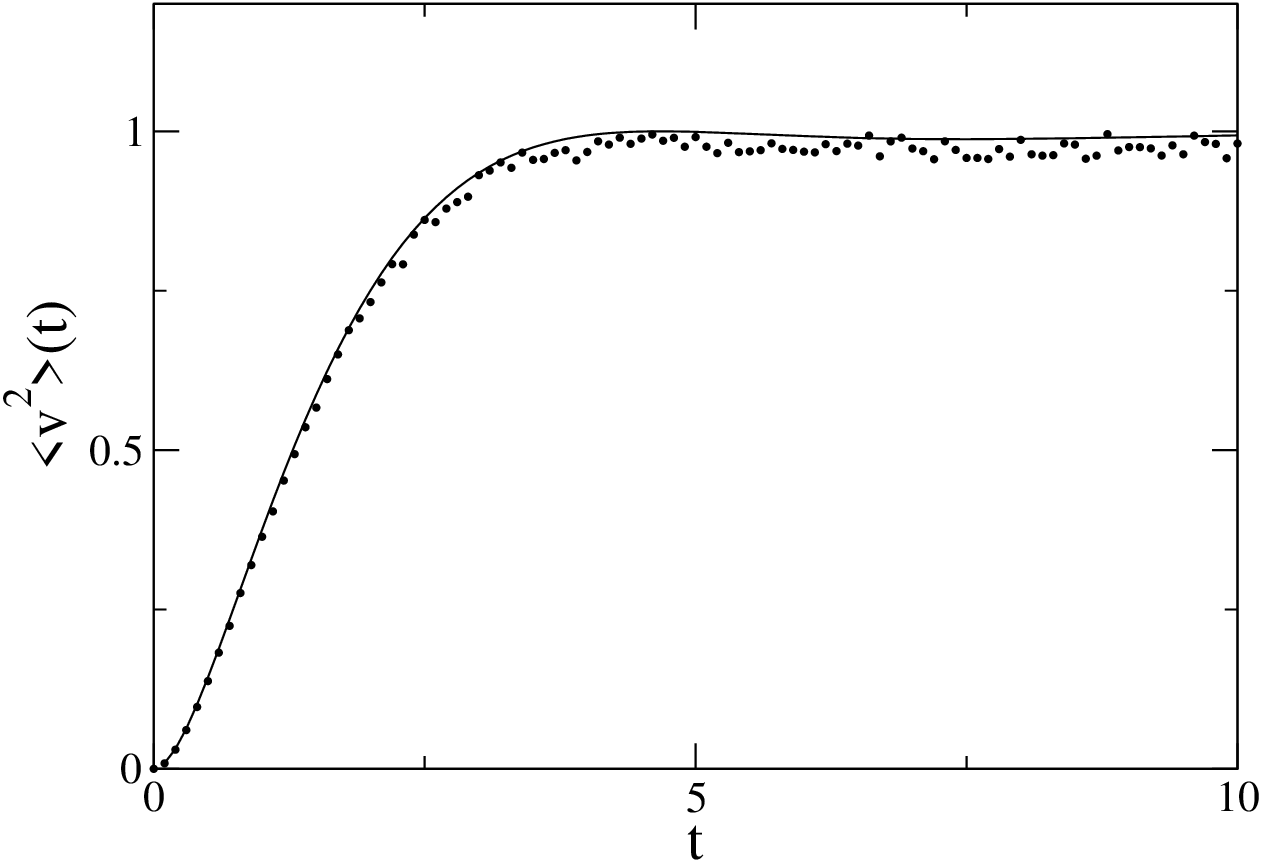}
\caption{The velocity variance calculated from Eq.(\ref{v2a}) (solid line) and
by numerical simulation from Eq.(\ref{solv1}) (dots). We assumed:
${\bf v}(0)=0$, $T=1$ and $m=1$.}
\end{figure}
over the stationary distribution of the random force (\ref{psta}). Fig.2
presents this quantity, calculated with the initial condition ${\bf v}(0)=0$,
for $T=1$ and $m=1$. On the other hand, the velocity variance can be derived
analytically from the Eq.(\ref{solv1}); the expression for
$\langle{\bf v}^2(t)\rangle$ involves only the second moment of the noise:
\begin{equation}
\label{v2a}
\langle{\bf v}^2\rangle(t)=m^{-2}\int_0^t\int_0^t R(\tau)\,R(\tau')\,
{\cal C}(|\tau-\tau'|)\,d\tau d\tau'.
\end{equation}
The velocity variance appears to be independent of a specific noise model
and then analytical and numerical results should coincide.
Indeed, Fig.2 demonstrates very good agreement of both results; they
indicate the relaxation to the thermal equilibrium
$\langle{\bf v}^2\rangle=k_BT/m$ which is apparently reached at about
$t=4$ \cite{uwa}.

\begin{figure}[htb]
\includegraphics[width=8.5cm]{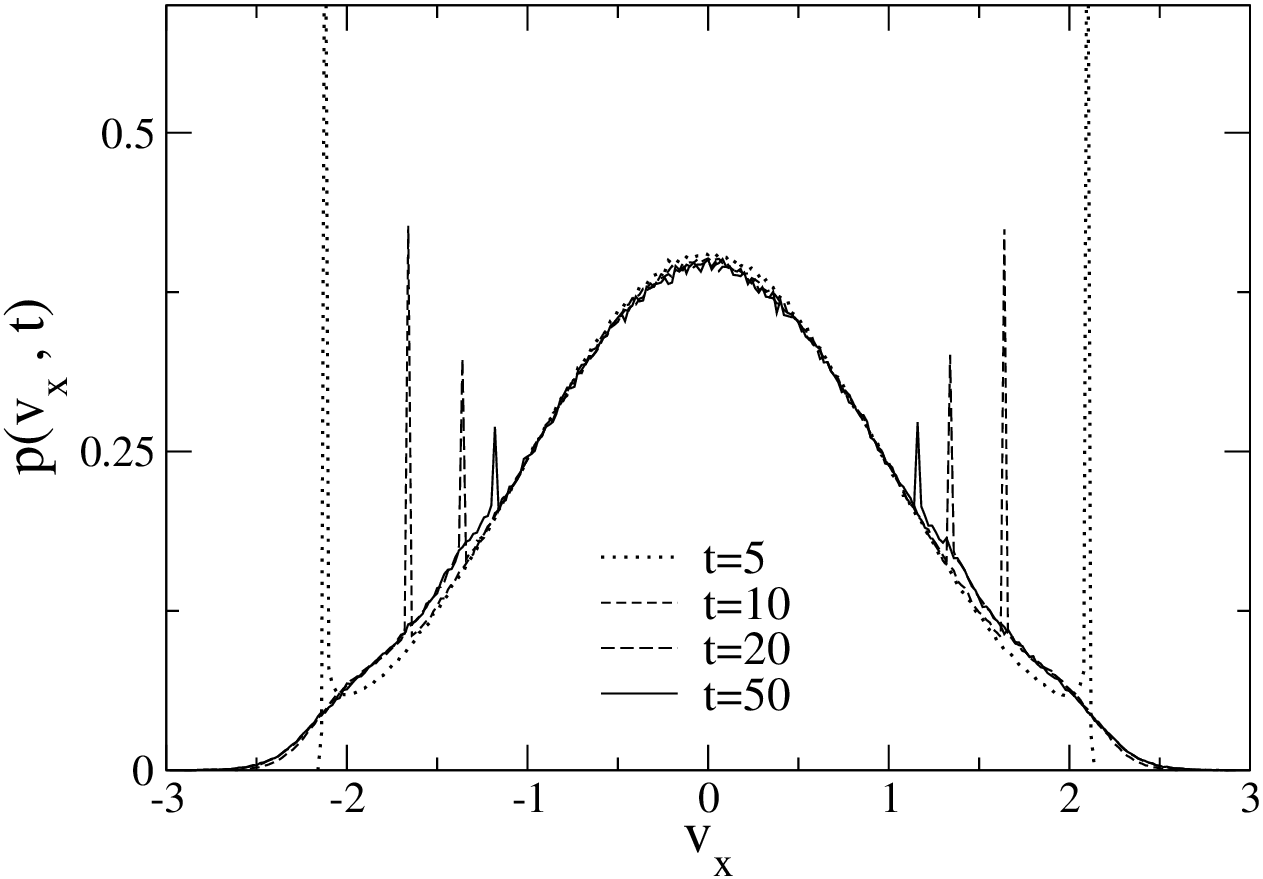}
\caption{Time evolution of the probability density distribution of the first
velocity component $v_x$. The stochastic ensemble consists of $5 \cdot
10^6$ trajectories for each time.}
\end{figure}

In a similar way, utilizing Eq.(\ref{solv1}), 
we can determine a density distribution
$p({\bf v},t)$ which means a probability that the velocity of the Brownian
particle is in the interval $({\bf v},{\bf v}+d{\bf v})$. Fig.3 presents this
distribution, corresponding to the first velocity component $v_x$, for large
times. The central part of the distribution is equilibrated already at $t=5$
but tails are not yet developed; 
they terminate with high and narrow peaks which
originate from the initial condition $p(v_x,0)=\delta(v_x)$. At short
times (not shown in the figure) the peaks are still higher and expand gradually
with time from a vicinity of the point $v_x=0$. Full relaxation of the tails --
which fall off faster than the Gaussian -- to the stationary distribution is
achieved at $t=20$. Nevertheless, the memory about the initial condition is
preserved for a very long time. The distribution of the second velocity
component $v_y$, presented in Fig.4, looks different; the width is much
smaller and the tails show the exponential shape. A complete relaxation to
stationary distribution is reached already at $t=10$. The difference between
the distributions for both velocity components follows from anisotropy of the
function $\nu(\phi)$: there are no infinite waiting times corresponding to the
motion in the $y$ direction.
\begin{figure}[htb]
\includegraphics[width=8cm]{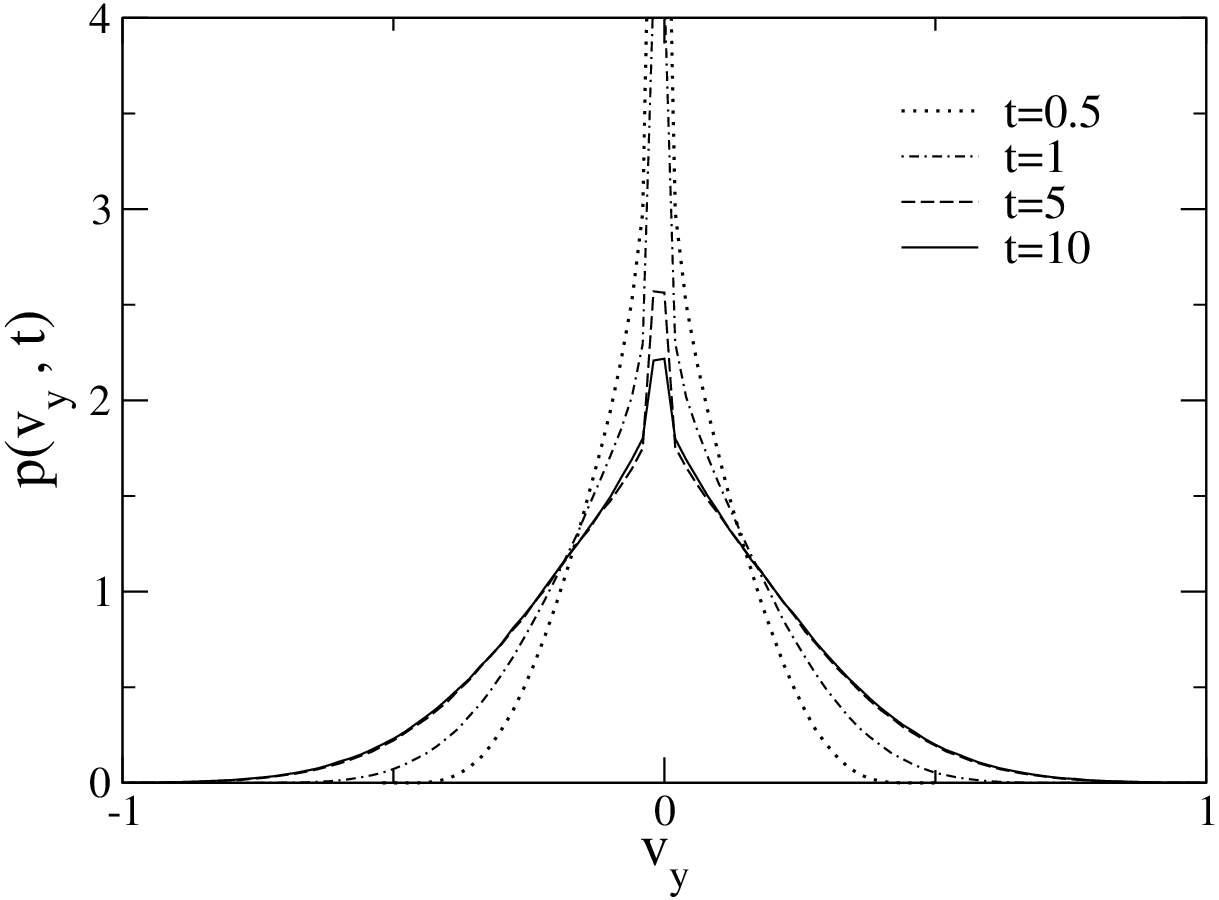}
\caption{The same as Fig.3 except for the second velocity component $v_y$.}
\end{figure}

The energy spectrum of the Brownian particles deviate considerably from the
Maxwellian distribution. Fig.5 presents the time evolution of the probability
density distribution of the energy $E=0.5(v_x^2+v_y^2)$. At small values of
the energy the curves have a cusp, whereas the tail of the distribution
corresponding to the equilibrium state can be parameterized by the function
$0.5\exp(-0.5E^2)~~(E>2)$. It is interesting that the probability density
function which characterizes the transport dynamics in the framework of the
continuous-time random walk, predicts a similar cusp for the subdiffusive
motion \cite{met}.
\begin{figure}[htb]
\includegraphics[width=8.5cm]{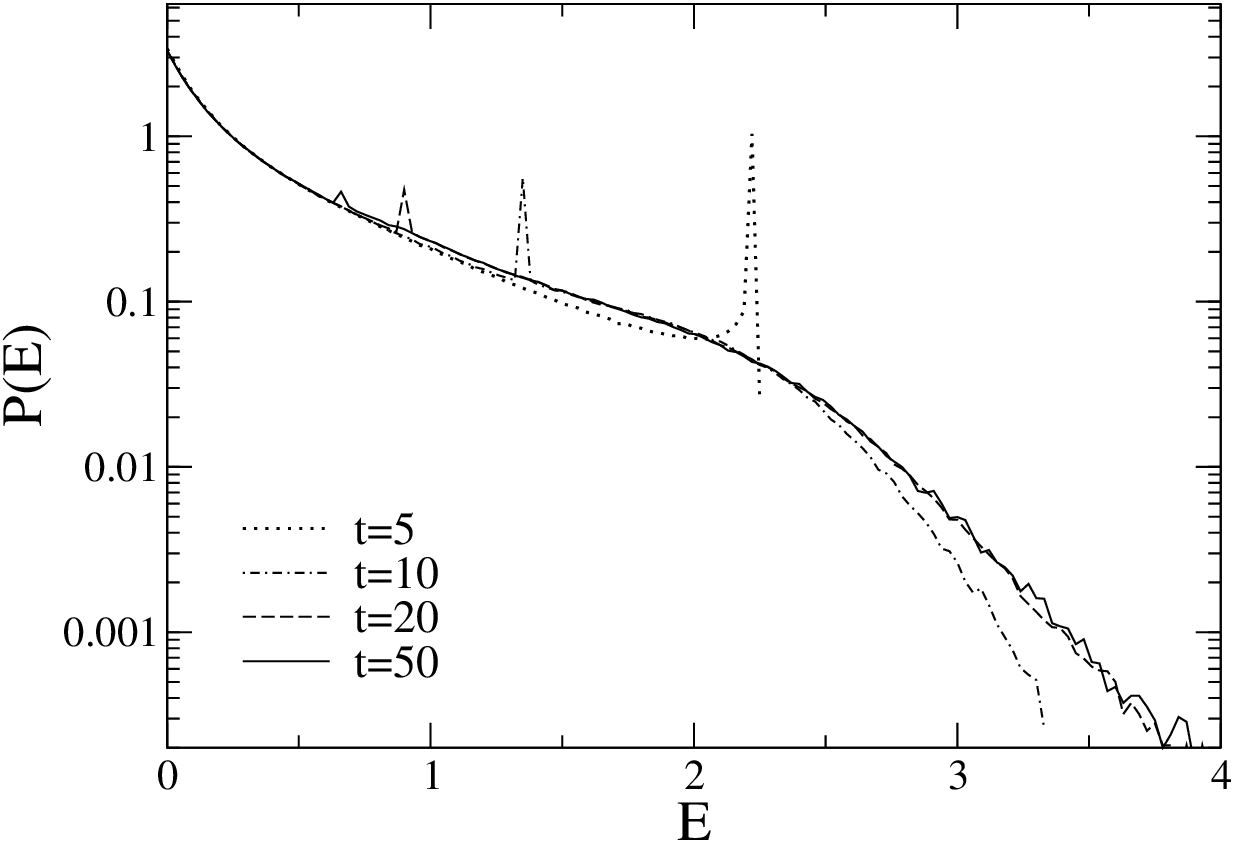}
\caption{Time evolution of the probability density distribution of the energy
$E=0.5(v_x^2+v_y^2)$.}
\end{figure}

\section{Summary and discussion}

The jumping process presented in this paper is characterized by
the jump size probability distribution and the waiting time
distribution which are correlated. The jumping rate depends on the
process value which is kept constant between consecutive jumps.
The process is Markovian and stationary, the corresponding master
equation possesses a nontrivial time-independent solution which is
completely determined by the jumping rate and which does not
depend on the jump size distribution. We have studied the process
in its two-dimensional version for jumps which do not change the
norm of the process value. The expression for ACF with power-law
tails has been derived. We have demonstrated that it is possible
to construct in a simple way a process which is a $1/f-$like noise.
\newline Despite the fact that the waiting time distribution is 
exponential, the intervals of constant process values can be very
long and actually algebraically distributed. This conclusion is
not surprising because the mean value of the exponential distribution
is also a stochastic variable. Then the
existence of long tails of the waiting time distribution does not
rule out a relaxation to the equilibrium.
\newline The procedure described in the paper allows us to construct stochastic
trajectories corresponding to a wide class of power law ACF
in a simple manner. Therefore it can serve as a model of physical
phenomena and can be used as a stochastic force in the generalized Langevin
equation. We have solved this equation for an exemplary form of ACF,
$\sim 1/t$, utilizing our process. Since waiting times are correlated with the
direction of the noise vector, the resulting velocity distribution exhibits a
strong anisotropy. The distribution of the first component, corresponding
to long waiting times, has rapidly falling tails
and indicates an extremely long
memory about the initial condition, despite the fact that the comprehensive
shape of the distribution equilibrates relatively fast. On the other hand,
the tails of the distribution corresponding to the second component
coincide with the standard Gaussian.
\newline The tail of ACF is determined predominantly by the long waiting times
and then only one component of the process value is crucial for its shape.
Therefore, this component can constitute a one-dimensional counterpart of our
two-dimensional jumping process which still has the power-law ACF. This remark 
is important if one requires a model of noise possessing 
an arbitrary dimensionality.

\end{document}